\documentclass{article}
\usepackage[utf8]{inputenc}
\usepackage[a4paper, total={5.5in, 9.4in}]{geometry}
\usepackage{chemformula}
\usepackage{graphicx} 
\usepackage{amsmath} 
\usepackage{amssymb}
\usepackage{amsthm}
\usepackage[utf8]{inputenc}
\usepackage[titles]{tocloft}
\usepackage{algorithm}
\usepackage{algorithmic}
\usepackage[newcommands]{ragged2e}
\usepackage[font=small,labelfont=bf,format=plain]{caption}
\DeclareCaptionJustification{justified}{\justifying}
\usepackage{setspace}
\usepackage{hyperref}
\usepackage{appendix}

\usepackage{comment}
\usepackage{copyrightbox}
\usepackage{enumitem}
\newlist{steps}{enumerate}{1}
\setlist[steps, 1]{label = \emph{Step} \arabic*.}
\usepackage{listings}
\usepackage{xcolor}
\usepackage{emptypage}
\usepackage{soulutf8}
\usepackage{extarrows}
\usepackage{csquotes}
\usepackage[T1]{fontenc}
\usepackage{authblk}
\numberwithin{equation}{section}
\newtheorem{theorem}{Theorem}[section]
\newtheorem{proposition}[theorem]{Proposition}

\newtheorem{corollary}[theorem]{Corollary}

\theoremstyle{definition}
\newtheorem{rem}[theorem]{Remark}

\newtheorem{definition}[theorem]{Definition}
\newtheorem{example}[theorem]{Example}

\definecolor{part}{HTML}{771155}
\definecolor{chap}{HTML}{AA4488}

\newcommand{\X}{X}
\newcommand{\Y}{Y}

\makeatletter
\renewcommand{\@biblabel}[1]{\quad#1.}
\makeatother
\title{Complete mathematical characterization of two simple toggle-switch biological systems}

\author[1]{Xavier Richard}
\author[2]{Benoît Richard}
\author[1]{Christian Mazza}
\author[3]{Jan Roelof van der Meer}
\affil[1]{Department of Mathematics, University of Fribourg, 1700 Fribourg, Switzerland}
\affil[2]{Department of Physics, University of Fribourg, 1700 Fribourg, Switzerland}
\affil[3]{Department of Fundamental Microbiology, University of Lausanne, 1015 Lausanne, Switzerland}
\date{}
\begin{document}
\maketitle
\justifying
\begin{abstract}
    Prokaryotic gene expression is dynamic and noisy, and can lead to phenotypic variations. Previous work has shown that the selection of such variation can be modeled through basic reaction networks described by O.D.E. displaying a bistable behavior. While previous mathematical studies have shown that mono- or bistable behavior depends on the rate of the reactions in the system, no analytical solution of the curves delineating the actual parameter conditions that result in mono or bistability has so far been provided.
    In this work we provide the first explicit analytical solution for the boundary curve that separates the parameter space defining domains where double positive and double negative feedback loops become bistable.
\end{abstract}

\section{Introduction}

Prokaryotic gene expression is dynamic and, as a result of noisy components and interactions, will lead to variation both in time and among individual cells \cite{Eldar2010, Balzsi2011, Pedraza2005, Thattai2004}. Gene expression variation will thus lead to phenotypic variation; the level of variation being different for individual networks or promoters \cite{Kussell2005}, with variability being a selectable trait.

In some cases gene expression networks not only lead to variation around a single mean phenotype, but can lead to two (or more) stable phenotypes - mostly resulting in individual cells displaying either the one or the other phenotype \cite{Ferrell2002, Dubnau2006}. Importantly, such bistable states are an epigenetic result of the network functioning and do not involve modifications or mutations on the DNA \cite{Balzsi2011, Kussell2005}. Bistable phenotypes may endure for a particular time in individual cells and their off-spring, or erode over time as a result of cell division or other mechanism, after which the ground state of the network reappears. 
Experimental bistable states have been modeled and produced by engineering of so-called toggle switches; two repressors under mutual control \cite{Gardner}. 
More general and conceptual frameworks of bistable networks were formulated in various papers such as by Ferrell \cite{Ferrell2002} and Balaszi \cite{Balzsi2011}, whereas the mathematics for bistable switches was described in detail by Tiwari \cite{Tiwari}, Jaruszewiz and others \cite{Jaruszewicz}.
Both, double positive feedforward ('inducible') and  double negative ('repressable') networks lead to bistable dynamic behaviour,  as illustrated in Figure \ref{double.1}.  Both types of networks occur in natural as well as engineered prokaryotic settings, such as competence formation in \textit{Bacillus subtilis} \cite{Dubnau2006}, the integrative and conjugative element ICEclc transfer competence in \textit{Pseudomonas putida} \cite{Carraro2020}, or the phage lambda lysogeny/lytic phase decision \cite{Ptashne}, and have been specifically modeled mathematically \cite{Suel2007, Bednarz2014, Arkin1998,Mazza2014}. 
Understanding the network configurations leading to bistable phenotypes had led to a number of mathematical approaches, which mostly start from chemical reaction networks, assuming mass-action kinetics \cite{Conradi, Mackey,SiegalGaskins2011,Siegal2}, and then derive ODE models \cite{Siegal2} or species-reaction graphs \cite{Craciun} to describe the network interactions and reaction direction or dynamics. Although this can lead to definition of the inherent capacity of a network to produce bistability \cite{Craciun}, delineating the actual parameter conditions that result in bistable behaviour is a more difficult mathematical problem \cite{Conradi}.
As examples, Conradi and coworkers \cite{Conradi} used polynomial analysis to define multistationarity in networks and bifurcation criteria to reject non-bistable situations or solutions, whereas or Siegal-Gaskins et al. \cite{SiegalGaskins2011, Siegal2}  discriminated bistable circuit behaviour on the basis of Sturm's theorem deriving exact analytical expressions. Both note that polynomial analysis cannot deduce the stability of the multistationary equilibria, or that not a single polynomial may describe the equilibrium state. Although such generic  reaction networks are the building blocks of many cellular reaction networks, to the best of our knowledge, no analytical solutions exist for the boundary solution between mono- and bistable regions in the parameter space of bistable networks for O.D.E. based mathematical models.
Our main result is an explicit analytical solution that is relevant for experimental and modelling research in systems biology.


\section{Chemical reaction network}
We start by defining a \emph{multimerization process} which represents the fact that $n$ factors of a chemical species $A$ combine to form a complex $nA$. This process is described by the following chemical reaction:
\begin{equation*}
    \ch{$\underbrace{A+...+A}_{n \text{ times}}$ -> $nA$}
\end{equation*}
Frequently, biological interaction with operators (such as e.g., DNA binding proteins on DNA) may need prior multimerization of $n$ factors $A$ or $m$ factors $B$. However, as described by Mazza and Benaï \cite{Mazza2014}, the multimerization reaction is fast compared to the other reactions in the networks we consider below. For simplicity, therefore, we neglect the multimerization reaction, and only consider the production and degradation reactions of both factors as critical for the binding process.
\label{Chemical reaction network}
\subsection{Double positive and double negative feedback loop}
\begin{figure}[h!]
    \centering
    \includegraphics[width=0.95\linewidth]{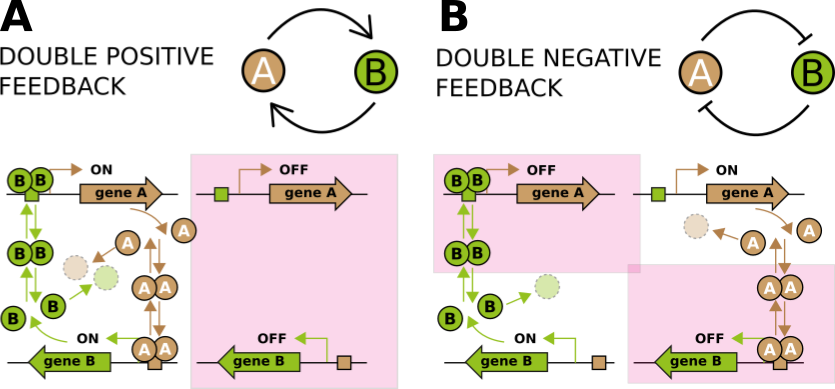}
    \captionof{figure}{Schematic representation of a biological double positive and negative feedback loop, formed by transcription factor dimer interactions on reciprocal promoter elements.  Transcription activator $A$ is expressed from gene $A$, and its dimer binds to the operator for the gene coding for transcription activator $B$. We assume that binding turns ON the promoter of gene $B$, leading to the production of factor $B$. $B$ dimerizes and binds to the operator of the gene for factor $A$, controlling reciprocal production of factor $A$. In case of a double negative feedback loop, both promoters are ON when they are unbound, leading to production of the reciprocal transcriptional repressors $A$ and $B$. Factor $A$ dimers bind the operator of $B$ and turn its promoter OFF, which blocks the production of $B$. Similarly, dimers of factor $B$ bind the $A$-operator and turn production of $A$ OFF.}
  \label{double.1}
\end{figure}

Consider a double positive feedback loop between the regulatory factors $A$ and $B$ i.e. $A$ promoting the formation of $B$ and $B$ promoting the formation of $A$, (Figure \ref{double.1} \textbf{A}) as the following chemical reaction network:
\begin{align}
\label{chemical reaction network double positive}
\begin{aligned}
  \ch{$mB$ + $\mathcal{O}^A$ &<=>[ $k_f^A$ ][ $k_b^A$ ] $\mathcal{O}^A_{mB}$ }\\
  \ch{$\mathcal{O}^A_{mB}$  &->[ $\mu^A$ ] $A+\mathcal{O}^A_{mB}$}\\
  \ch{$A$  &->[ $\nu^A$ ] $\emptyset$}
\end{aligned}
&&
\begin{aligned}
    \ch{$nA$ + $\mathcal{O}^B$ &<=>[ $k_f^B$ ][ $k_b^B$ ] $\mathcal{O}^B_{nA}$ }\\
  \ch{$\mathcal{O}^B_{nA}$  &->[ $\mu^B$ ] $B+\mathcal{O}^B_{nA}$}\\
  \ch{$B$  &->[ $\nu^B$ ] $\emptyset$}.
\end{aligned}
\end{align}
We use $\kappa^A_f$ and $\kappa^A_b$ to denote the binding and unbinding rates, respectively, of factor $B$ to the operator DNA controlling the production of $A$ from its respective gene. ${\mathcal{O}^A}$ represents the empty $A$-operator in absence of binding of $B$. This corresponds to the situation where the gene for the factor $A$ is not expressed. ${\mathcal{O}^A_{mB}}$ represents the situation where $m$ factors $B$ are bound to the $A$-operator. In this situation gene $A$ is turned on, yielding $A$ with a production rate equal to $\mu^A$. We consider that produced factor $A$ is degraded with a rate of $\nu^A$. The same notations are valid for the production of factor $B$ from its B-gene as a result of interaction of $n$ $A$-factors with the $B$-operator, and its subsequent degradation.

Similarly, we describe a double negative feedback loop between regulatory factors $A$ and $B$, for which $B$ inhibits the formation of $A$, and $A$ inhibits formation of $B$ (Figure \ref{double.1} \textbf{B}). In this case, the chemical reaction network is described as:
\begin{align}
\label{chemical reaction network double negative}
    \begin{aligned}
        \ch{$mB$ + $\mathcal{O}^A$ &<=>[ $k_f^A$ ][ $k_b^A$ ] $\mathcal{O}^A_{mB}$ }\\
        \ch{$\mathcal{O}^A$  &->[ $\mu^A$ ] $A+\mathcal{O}^A$}\\
        \ch{$A$  &->[ $\nu^A$ ] $\emptyset$}
    \end{aligned}
    &&
    \begin{aligned}
        \ch{$nA$ + $\mathcal{O}^B$ &<=>[ $k_f^B$ ][ $k_b^B$ ] $\mathcal{O}^B_{nA}$ }\\
      \ch{$\mathcal{O}^B$  &->[ $\mu^B$ ] $B+\mathcal{O}^B$}\\
      \ch{$B$  &->[ $\nu^B$ ] $\emptyset$}
    \end{aligned}
\end{align}
with the same notations as those for the case of the double positive feedback loop of Equation (\ref{chemical reaction network double positive}). In case of the double negative feedback, binding of $m$ factors $B$ to the operator $\mathcal{O}^A$ blocks the production of factor $A$. When the operator $\mathcal{O}^A$ is free of factor $B$, it produces factor $A$ at rate $\mu^A$. Similarly, production of factor $B$ is inhibited when the operator $\mathcal{O}^B$ is bound by $n$ factors $A$, and $B$ is produced at a rate equal to $\mu^B$ when the operator $\mathcal{O}^B$ is unbound (Figure \ref{double.1}).
\section{Main results}
The main question of this work is to determine for which conditions the systems can be bistable. 
\\

First, we use mass action kinetics defined to rewrite the systems of chemical equations \ref{chemical reaction network double positive} and \ref{chemical reaction network double negative} as systems of differential equations. At equilibrium, the double positive feedback corresponds to the system
\begin{align}
    \label{Double positive feedback simplified}
      \begin{cases}
      \dfrac{c_1 \Y^m}{1 + \Y^m} - \X = 0
      \\
      \\
      \dfrac{c_2 \X^n}{1 + \X^n} - \Y = 0
    \end{cases}
\end{align}
and the double negative feedback to the system
\begin{align}
    \label{Double negative feedback simplified}
      \begin{cases}
      \dfrac{c_1}{1 + \Y^m} - \X = 0
      \\
      \\
      \dfrac{c_2}{1 + \X^n} - \Y = 0.
    \end{cases}
\end{align}
For these two systems, the parameters $c_1$ and $c_2$ are constants, while the variables $X$ and $Y$ depend on the concentrations of $A$ and $B$ respectively. A precise derivation of this result is given in Appendix \ref{appendix: from chemical to ode}.

\begin{proposition}
    \label{prop:3steady}
    The systems \ref{Double positive feedback simplified} and \ref{Double negative feedback simplified} are bistable if and only if they have three steady states.
    \begin{proof}
        A proof of this proposition is given in Section \ref{sec:proofs}
    \end{proof}
\end{proposition}
The next step is to mathematically describe the phase diagram of a double positive and double negative feedback loop system with respect to the parameters $c_1$ and $c_2$, the two possible phases of such system being monostable or bistable. Proposition \ref{prop:3steady} states that a monostable phase corresponds to a system having one or two steady states, whereas a bistable phase arises in a system with exactly three steady states. To decide on the phase type of a system we need to find the mathematical conditions describing its number of steady states (corresponding to the phase transitions), and use this to characterize the critical region separating mono- and bistable phases. This critical region appears to be a curve and, we therefore refer to it as the critical curve. The critical curves for the double positive and the double negative feedback loop are given by the following theorem.
\begin{theorem}
    \label{th:curves}
    \hfill
\begin{itemize}
\item The critical curve for the double positive feedback loop (system \ref{Double positive feedback simplified}) is given by the following system of parametric equations
\begin{align}
    \label{Boundary curve double positive feedback}
    \begin{cases}
        c_1(\X) = \dfrac{mn \X}{mn - 1 - \X^n}\\
        \\
        c_2(\X) = \left(1 + \dfrac{1}{\X^n}\right) \left(\dfrac{mn}{1 + \X^n} - 1 \right)^{\frac{1}{m}}.
    \end{cases}
\end{align}
\item The critical curve for the double negative feedback loop (system \ref{Double negative feedback simplified}) is given by the following system of parametric equations
\begin{align}
    \label{Boundary curve double negative feedback}
    \begin{cases}
        c_1(\X) &=  \dfrac{m n \X^{n + 1}}{(m n - 1) \X^n - 1}\\
        \\
        c_2(\X) &= (1 + \X^n) \left(\dfrac{1 + \X^n}{(m n - 1) \X^n - 1}\right)^{\frac{1}{m}}.
    \end{cases}
\end{align}
\end{itemize}
\begin{proof}
    A proof of this theorem is given in Section \ref{sec:proofs}
\end{proof}
\end{theorem}

\begin{corollary}
    \label{cor:2regions}
    The critical curves given by Theorem \ref{th:curves} separate their respective phase diagram into exactly two regions.
    \begin{proof}
        A proof of this corollary is given in Section \ref{sec:proofs}
    \end{proof}
\end{corollary}

Now that we have found and characterized the critical curve separating the mono- of the bistable phases in a double positive and double negative feedback loop, we need to determine the number of steady states in each phase. 

\begin{figure}[h!]
    \centering
    \includegraphics[width=0.95\linewidth]{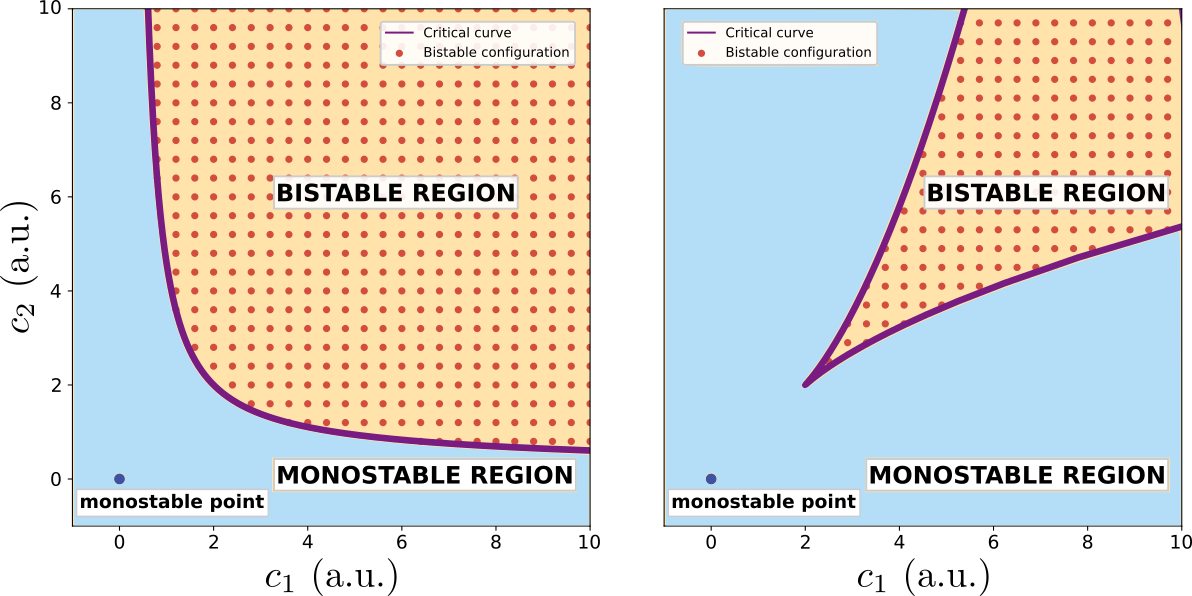}
    \caption{Diagram of the mono and bistable phases for the double positive (\textbf{left}) and double negative ({\textbf{right}}) feedback loop systems, for $m = n = 2$. The critical curves are computed analytically from Theorem \ref{th:curves}. The dots are point for which the bistability of the system has been rigorously verified numerically using interval arithmetic, as described in Appendix \ref{double:simulation}.
    }
    \label{Figure5.7}
\end{figure}

\begin{corollary}
    \label{cor:0mono}
    The region containing the configuration $c_1 = c_2 = 0$ is monostable, while the other region is bistable. Moreover, the critical curve belongs to the monostable region.
    \begin{proof}
        For the case when $c_1 = c_2 = 0$, the two stationary curves (Equations \eqref{Double positive feedback simplified} and \eqref{Double negative feedback simplified}) are straight lines and therefore only one steady state is present. This indicates that the phase containing the $(0, 0)$ point is monostable (Figure \ref{Figure5.7}). In the other phase, on the other side of the critical region, we found parameter combinations permitting three steady states (Figure \ref{bist_3fig} \textbf{B}). This other phase is therefore bistable. Finally, the parameter combinations on the critical curve admit either one or two solutions, thus the critical curve is part of the monostable phase according to the Proposition \ref{prop:3steady}.
    \end{proof}
\end{corollary}

Corollary \ref{cor:0mono} fully determines the bistability phase diagram for all the double positive and double negative feedback loop systems in terms of the parameters $c_1$ and $c_2$. Those results are illustrated in Figure \ref{Figure5.7}. Numerical  verifications developed in Section \ref{double:simulation} confirm those results.

\section{Proofs}
\label{sec:proofs}
\subsection{Proof of proposition \ref{prop:3steady}}
The proof of Proposition \ref{prop:3steady} consists of three steps:
\begin{enumerate}
    \item The steady states of systems \ref{Double positive feedback simplified} and \ref{Double negative feedback simplified} are characterized in terms of the eigenvalues of their respective Jacobian matrix. We then show that the determinant of the Jacobian matrix is sufficient to give a complete characterization of the steady states.
    \item A geometrical interpretation of these steady states is given using phase plane analysis.
    \item The proposition is proved using the determinant and the considerations on the plane phase analysis.
\end{enumerate}
\subsubsection{Stability of steady states}
Equations \eqref{Double positive feedback simplified} and \eqref{Double negative feedback simplified} are valid under \emph{steady state} conditions, which are characterized by the fact that the concentrations of the components in the system do not change, since the time derivative of both variables is zero. However, steady states relevant for biological systems can be divided in those which are \emph{stable} and those which are \emph{unstable}. Remember that a steady state is stable when it always recovers after a small perturbation and returns to its previous state. To determine the stability of a steady state of a system of two equations with two variables
\begin{equation}
\label{general equations}
\begin{cases}
f(\X, \Y)=0
\\
g(\X, \Y)=0,
\end{cases}
\end{equation}
it is standard to look at the eigenvalues of its Jacobian matrix $J$. They are given by
\begin{align}
\label{eigenvalues}
\begin{aligned}
	\lambda_{1} = \frac{1}{2} \left(T + \sqrt[]{T^2-4\delta}\right)
\end{aligned}
&&
\begin{aligned}
    \lambda_{2} = \frac{1}{2}\left(T - \sqrt[]{T^2-4\delta}\right)
\end{aligned}
\end{align}
where $T$ is the trace of the matrix $J$ and $\delta$ is its determinant. If both eigenvalues $\lambda_1$, $\lambda_2$ are negative, the steady state solution is stable. If the eigenvalues are either both positive or of different sign, the steady state is unstable. If at least one of the eigenvalues is zero, the nature of the steady state solution (stable or unstable) cannot be determined from $\lambda_1$ and $\lambda_2$ \cite{Iooss1990,Izhikevich2007,Strogatz1994}.
\\

We can now write explicitly the Jacobian matrix of the systems \eqref{Double positive feedback simplified} and \eqref{Double negative feedback simplified}. However, we first introduce the concept of Hill function to simplify the notations.

\subsubsection*{Hill functions}
\begin{figure}[h!]
  \centering
    \includegraphics[width=0.9\linewidth]{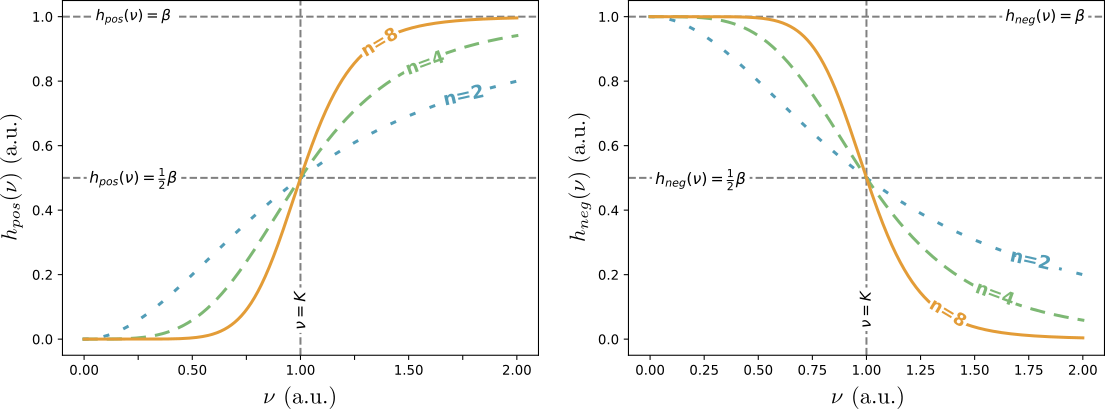}
   \captionof{figure}{Examples of positive (\textbf{left}) and negative (\textbf{right}) Hill functions. Parameters are $K=1$, $\beta=1$ and the Hill coefficient $n$ varies according to the legends.}
 \label{hill_function}
\end{figure}

Hill functions, firstly introduced by Hill \cite{Hill1910}, and defined in more detail in e.g. \cite{Alon2006,Mazza2014} are typically used to model a system when a production $\mu$ of a component $B$ depends on the concentration $\nu$ of a component $A$. A \emph{positive Hill function} which is defined by 
\begin{equation*}
    h_{pos}(\nu) = \beta\frac{\nu^{n}}{K^{n}+\nu^{n}}
\end{equation*}
models an activator, an increase of the concentration $\nu$ increase the production of $B$. On the contrary, a \emph{negative Hill function} defined by 
\begin{equation*}
    h_{neg}(\nu) = \beta\frac{K^{n}}{K^{n}+\nu^{n}}
\end{equation*}
models a repressor, an increase of the $\nu$ concentration decreases the production of $B$. Hill functions depend on three parameters $K$, $\beta$ and $n$. The \emph{activation coefficient} $K$ is a positive constant representing the half-saturation concentration; i.e., when $h_{pos}(K)\cdot 0.5\cdot\beta$ for a positive Hill function and $h_{neg}(K)\cdot 0.5\cdot\beta$ for a negative Hill function. The variable $\beta$ corresponds to the maximal production rate and finally the \emph{Hill coefficient} $n$ gives the steepness of the curve. 
The larger $n$ is, the steeper the curve will be, looking more and more like a step function. Examples of positive and negative Hill functions are illustrated in Figure \ref{hill_function}.

We can rewrite the double positive feedback loop system of Equation \eqref{Double positive feedback simplified} as
\begin{equation}
\label{Double positive with feedback rewritten}
\begin{cases}
	F_{pos}(\Y) - \X = 0\\
    G_{pos}(\X) - \Y = 0,
\end{cases}    
\end{equation}
where
\begin{equation}
\label{Definition of positive Hill functions}
\begin{cases}
    F_{pos}(\Y) = \dfrac{c_1 \Y^m}{1 + \Y^m}
    \\
    \\
    G_{pos}(\X) = \dfrac{c_2 \X^n}{1 + \X^n}
\end{cases}
\end{equation}
are positive Hill functions. 
\\
The Jacobian of the system \eqref{Double positive with feedback rewritten} transforms to
\begin{equation}
	J_{pos} = \begin{pmatrix} -1 & \dfrac{d F_{pos}(B)}{d B} \\ \dfrac{d G_{pos}(A)}{d A} & -1 \end{pmatrix}
	=\begin{pmatrix} -1 & \dfrac{c_1mY^{m-1}}{\left(1+Y^m\right)^2} \\ \dfrac{c_2nX^{n-1}}{\left(1+X^n\right)^2} & -1 \end{pmatrix}
\end{equation}
with trace
\begin{equation}
	\label{Trace of the Jacobi matrix}
	T = -2
\end{equation}
and determinant
\begin{equation}
	\label{Determinant of the Jacobi matrix}
	\delta = 1 - \dfrac{d F_{pos}(\Y)}{d\Y}\dfrac{d G_{pos}(\X)}{d\X}.
\end{equation}
In the case of a double negative feedback loop, we can rewrite the system of Equation \eqref{Double negative feedback simplified} as
\begin{equation}
\label{Double negative with feedback rewritten}
\begin{cases}
	F_{neg}(\Y) - \X = c_1 - F_{pos}(\Y) - \X = 0\\
    G_{neg}(\X) - \Y = c_2 - G_{pos}(\X) - \Y = 0,
\end{cases}
\end{equation}
where
\begin{equation}
\label{Definition of negative Hill functions}
\begin{cases}
    F_{neg}(\Y) = \dfrac{c_1}{1 + \Y^m}
    \\
    \\
    G_{neg}(\X) = \dfrac{c_2}{1 + \X^n}
\end{cases}
\end{equation}
are negative Hill functions (see \cite{Alon2006,Mazza2014}).
To obtain Equation \eqref{Double negative with feedback rewritten} we have used the fact that
\begin{equation}
\begin{cases}
    F_{pos}(\Y) + F_{neg}(\Y) = c_1
    \\
    G_{pos}(\X) + G_{neg}(\X) = c_2.
\end{cases}
\end{equation}
The Jacobian matrix associated to the double negative feedback loop case is then
\begin{equation}
	J_{neg} = \begin{pmatrix} -1 & -\dfrac{d F_{pos}(\Y)}{d\Y} \\ -\dfrac{d G_{pos}(\X)}{d\X} & -1 \end{pmatrix},
\end{equation}
and therefore the trace $T$ and determinant $\delta$ of $J_{neg}$ have the same form as for the Jacobian matrix $J_{pos}$ in the double positive feedback case. As a consequence, its eigenvalues are also given by Equations \eqref{Trace of the Jacobi matrix} and \eqref{Determinant of the Jacobi matrix}. Note that the definitions of $\X$, $\Y$, $c_1$ and $c_2$ are different for the double positive and double negative feedback cases (Equations \eqref{Normalized parameters}, \eqref{c1c2pos}, \eqref{XYneg}, and \eqref{c1c2neg}).
\\

Since the trace $T$ of the Jacobian matrix is a constant, the stability of a steady state depends only on its determinant $\delta$. From Equation \eqref{eigenvalues} it follows that $\lambda_2$ is negative for all values of $\delta$, and that $\lambda_1$ is positive only if $\delta$ is negative. Therefore, the steady state is stable if $\delta > 0$ and unstable if $\delta < 0$. If $\delta = 0$, the stability of the steady state cannot be concluded. A geometrical interpretation of these observations is given in the next section.

\subsubsection{Phase plane analysis}
\label{Phase plane analysis}
\begin{figure}[h!]
    \centering
    \includegraphics[width=0.95\linewidth]{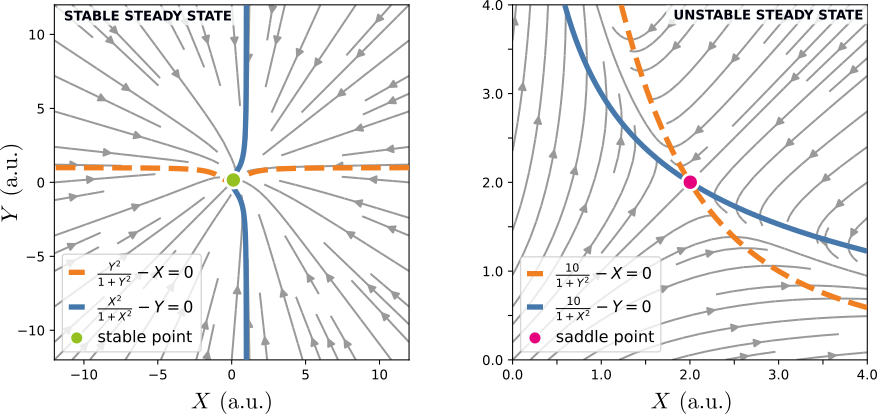}
    \caption{Example of appearance of stable and unstable steady state nodes as in the indicated functions. A stable node is characterized by a convergence of all streamlines (lines with arrows) toward it. An unstable steady state node is characterized by some streamlines passing through, but most others leading away. Orange and blue lines indicate the cases, where one of the two variables ($X$ or $Y$, as defined in the main text) is zero.}
  \label{Figure5.2}
\end{figure}

A \emph{phase plane} is a graphical representation of a system of differential equations. The goal of such representation is to indicate how the system evolves as a function of its parameters. Because the reduced systems of Equations \eqref{Double positive feedback simplified} and \eqref{Double negative feedback simplified} are deterministic, their evolution is uniquely determined by the initial state, and their different possible trajectories (often referred to as streamlines in the context of phase plane analysis) can be represented in the $\X\Y$ phase plane as directed curves. These trajectories only intersect at steady states (Figure \ref{Figure5.2}). When all streamlines are leading toward the same point, the corresponding steady state is stable, and a small perturbation would follow the streamlines back to the original steady state. In contrast, if some streamlines lead away from the steady state, a perturbation may push the system away from the steady state, making it an unstable steady state, as illustrated in Figure \ref{Figure5.2}. The points of the $\X\Y$ plane for which the derivative of $\X$ is zero define a phase line, the equation of which can be deduced from Equations \eqref{Double positive with feedback rewritten} or \eqref{Double negative with feedback rewritten}. The same is true for the region of the $\X\Y$ plane where $\Y$ is constant, defining two distinct phase lines whose intersections correspond to steady state as the derivatives of both $\X$ and $\Y$ are zero (Figure \ref{Figure5.2}).

Since the reasoning and description are similar for both the double positive feedback loop and the double negative feedback loop, we only present here the details for the double positive feedback. For the double positive feedback system of Equation \eqref{Double positive with feedback rewritten}, the two phase lines are described by 
\begin{equation}
\label{Stationary curves as functions}
\begin{cases}
	\X = F_{pos}(\Y) \\
    \Y = G_{pos}(\X),
\end{cases}
\end{equation}
where the first equation can be interpreted as representing the variable $\X$ as a function of $\Y$ and the second as representing $\Y$ as a function of $\X$. Because $F_{pos}$ is continuously differentiable, and its derivative is invertible for $\Y > 0$, we can use the inverse function theorem to locally invert it. We can express the inverse function as
\begin{equation}
	\Y = \widetilde{F}_{pos}(\X).
\end{equation}
The derivative of this new function $\widetilde{F}_{pos}(\X)$ is then related to the derivative of $F_{pos}(\X)$ by
\begin{equation}
	\frac{d \widetilde{F}_{pos}(\X)}{d \X} = \left(\frac{d F_{pos}(\Y)}{d \Y}\right)^{-1},
\end{equation}
This allows to rewrite the determinant $\delta$ of the Jacobian matrix in Equation \eqref{Determinant of the Jacobi matrix} as
\begin{equation}
	\label{Determinant of the Jacobi matrix with inverse function}
	\delta = 1 - \frac{G_{pos}'(\X)}{\widetilde{F}'_{pos}(\X)}
\end{equation}
where the prime denotes the derivative relative to $\X$. In the case of a double positive feedback loop the two phase lines always have positive derivatives for $\X \geq 0$; the region at $\X < 0$ would imply negative concentrations. 

\begin{figure}[h!]
    \centering
    \includegraphics[width=0.7\linewidth]{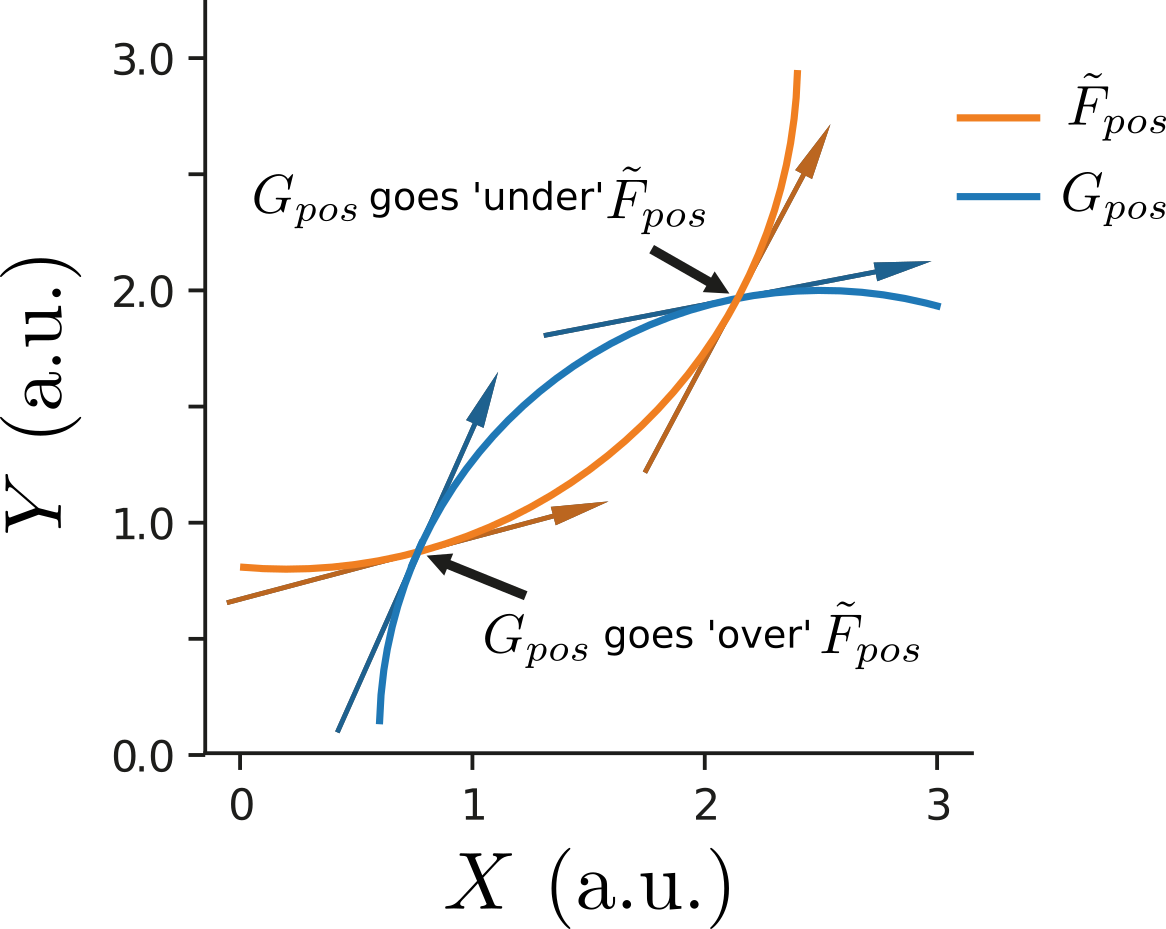}
    \caption{Illustration of intersections between two curves. Slopes at the intersection points are outlined by bold arrows. Arrows point in the direction of increasing $X$ (left to right) to clarify what it is understood by the terms \emph{going over} or \emph{going under}.}
  \label{Figure5.3}
\end{figure}

The condition for a stable node $\delta > 0$ at an intersection is, therefore, equivalent to $G_{pos}'(\X) < \widetilde{F}_{pos}'(\X)$, which can be interpreted as the curve $G_{pos}(\X)$ going from \emph{over} to \emph{under} the curve $\widetilde{F}_{pos}(\X)$. Similarly, an unstable steady state, characterized by $\delta < 0$, corresponds to the phase line $G_{pos}(\X)$ passing from below to \emph{over} the curve $\widetilde{F}_{pos}(\X)$. Both cases are illustrated in Figure \ref{Figure5.3}. Finally, in the edge case $\delta = 0$, $G_{pos}'(\X) = \widetilde{F}_{pos}'(\X)$, and the two curves are tangent.

\subsubsection{Bistability}
In the following, we show that the information given by the determinant $\delta$ of the Jacobian matrix is sufficient to determine the number of steady states in a system, and that this information is sufficient to conclude whether bistability (i.e., having two stable steady states) is possible in the chosen network configuration.

\begin{figure}[p!]
    \centering
    \includegraphics[width=0.95\linewidth]{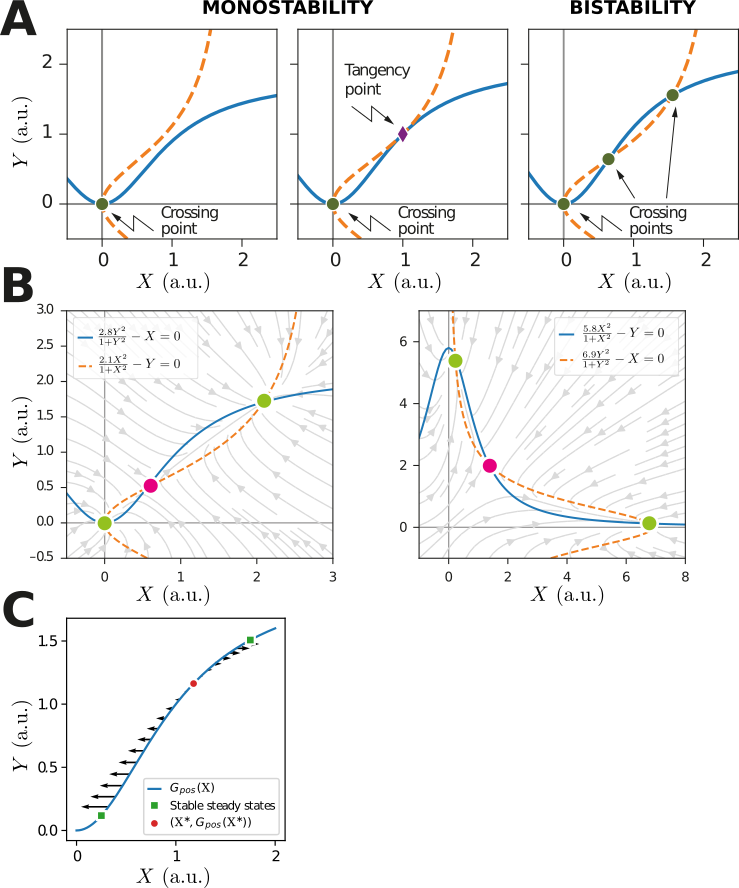}
    \caption{\textbf{A.} Change of the stationary curves for a double positive feedback loop with $m = n = 2$ when the parameters go from $c_1 = c_2 = 1.8$ (left, one intersection), to $c_1 = c_2 = 2$ (center, tangency point) and $c_1 = c_2 = 2.2$ (right, three intersections, occurrence of bistability). \textbf{B.} Phase plane analysis of a double positive and double negative feedback system, with parameters $c_1 = 2.8$ and $c_2 = 2.1$, and $c_1 = 6.9$ and $c_2 = 5.8$, respectively. Blue and orange lines indicate situations where one of the variables is zero. In both cases, there are two stable nodes (green circles) and one unstable node (magenta circle), $c_1$ and $c_2$ as defined in main text, and differently for double positive (Equation \eqref{c1c2pos}) or double negative system (Equation \eqref{c1c2neg}). \textbf{C.} Schematic representation of $\left(\frac{d\X}{dt}, \frac{d\Y}{dt}\right)$ as vectors along the steady curve $G_{pos}$ between two stable steady states. As explained in the text, this implies the presence of a third steady state $(X^*, G_{pos}(X^*))$.}
    \label{bist_3fig}
\end{figure}

First, in case that the curve $\widetilde{F}_{pos}(\X)$ is crossing the curve $G_{pos}(\X)$ only once, there is only one steady state (Figure \ref{bist_3fig} \textbf{A}). One steady state is insufficient to create true bistability. For bistability, $G_{pos}$ starts by going over $\widetilde{F}_{pos}$; the next intersection is necessarily in the opposite direction, $G_{pos}$ going under $\widetilde{F}_{pos}$, and the final intersection will be of the same type as the first one. As a consequence, both the first and the last point have a $\delta < 0$ and are thus stable, while the middle one is unstable with $\delta > 0$. Therefore, the system has three steady states (the maximum possible due to the geometry of the phase lines), (Figure \ref{bist_3fig} \textbf{A}) among which two stable states (Figure \ref{bist_3fig} \textbf{B}). The system is therefore called bistable.

Finally, in the case where the curves have two intersection points, one of them is tangent (Figure \ref{bist_3fig} \textbf{A}), and the following proposition prove that the system is not bistable. 
\begin{proposition}
Consider the two phase line deduced from equation \eqref{Double positive with feedback rewritten}. If the two curves have exactly
two intersection points then the associated system can not be bistable.
\begin{proof}
We prove that if the system admits only two steady states, that are both stable, we have a contradiction. Consider the double positive feedback case and the curve $G_{pos}(\X)$. On this curve we have, by definition $\frac{d\Y}{dt} = 0$. Moreover, near $\X_1$, we must have $\frac{d\X}{dt} > 0$ for $\X > \X_1$, since streamlines must go toward this steady state as it is stable. Similarly, near $\X_2$, we must have $\frac{d\X}{dt} < 0$ for $\X < \X_2$. Applying the intermediate value theorem to $\frac{d\X}{dt}$ on the steady curve $G_{pos}$, we find that there is $\X^* \in (\X_1, \X_2)$ such that $\frac{d\X}{dt} = 0$ at the point $(X^*, G_{pos}(X^*))$. Since this point is on the steady curve, we have both $\frac{d\X}{dt} = 0$ and $\frac{d\Y}{dt} = 0$. Therefore, the point $(\X^*, G_{pos}(\X^*))$ is a steady state as illustrated in Figure \ref{bist_3fig} \textbf{C}.

This implies that the system has at least three steady states, which contradicts our hypothesis. We therefore conclude that if two steady states exist, they cannot both be stable.
\end{proof}
\end{proposition}
We can therefore conclude that a system is bistable if and only if three steady states are present. Similar reasoning also applies for the double negative feedback loop (Figure \ref{bist_3fig} \textbf{B}).
\clearpage
\subsection{Proof of theorem \ref{th:curves}}

\subsubsection{Double positive feedback loop}
To determine the critical curve, we use the fact that a change in $c_1$ or $c_2$ is equivalent to smoothly deforming the stationary curves described by Equation \eqref{Stationary curves as functions}. As is illustrated in Figure \ref{bist_3fig} A, changing $c_1$ or $c_2$ leads to situations with different steady states, with a boundary state where the two stationary curves are tangent (Figure \ref{bist_3fig} A, middle panel). The condition when tangent stationary curves occur thus defines the transition from mono- to bistability. The necessary condition for stationary curves to be tangent is if and only if $\delta = 0$ which corresponds to the equation
\begin{equation}
    \label{proofth:d}
    1-\frac{c_1 c_2 n m X^{n-1} Y^{m-1}}{\left(1+X^n\right)^2 \left(1+Y^m\right)^2} = 0.
\end{equation}
This, together with Equation \eqref{Stationary curves as functions}, rewritten in terms of $c_1$ and $c_2$ as
\begin{align}
    \label{proofth:c1c2}
    \begin{cases}
        c_1 = \dfrac{X\left(1+Y^m\right)}{Y^m}\\
        \\
        c_2 = \dfrac{Y\left(1+X^n\right)}{X^n}
    \end{cases}
\end{align}
defines a critical curve. Substituting $c_1$ and $c_2$ in Equation \ref{proofth:d} gives
\begin{equation}
    \label{proofth:subs}
    1- \dfrac{nm}{\left(1+X^n\right) \left(1+Y^m\right)}=0
\end{equation}
which leads to
\begin{equation}
    \label{proofth:ym}
    \Y^m = \dfrac{nm -1 - \X^n}{1+\X^n}.
\end{equation}
Finally, by substituting Equation \eqref{proofth:ym} into \eqref{proofth:c1c2}, we obtain a system that can be parameterized in terms of $\X$ as
\begin{align*}
\begin{cases}
	c_1(\X) = \dfrac{mn \X}{mn - 1 - \X^n}\\
    \\
	c_2(\X) = \left(1 + \dfrac{1}{\X^n}\right) \left(\dfrac{mn}{1 + \X^n} - 1 \right)^{\frac{1}{m}}.
\end{cases}
\end{align*}

\subsubsection{Double negative feedback loop}

Using a similar procedure as for the double positive feedback loop, we obtain the critical curve for the double negative feedback loop given by \eqref{Boundary curve double negative feedback}.
\begin{rem}
    The critical curves \ref{Boundary curve double positive feedback} and \ref{Boundary curve double negative feedback} can be similarly parameterized with respect to $\Y$.
\end{rem}

\subsection{Proof of corollary \ref{cor:2regions}}
\subsubsection{Double positive feedback loop}
 The critical curve \ref{Boundary curve double positive feedback} is well-defined and continuous for
\begin{align*}
	0 < \X < \X_{sup},
\end{align*}
where $X_{sup} = (mn -1)^{\frac{1}{n}}$ is the maximal value that $X$ can take, as $X$ must be positive, and values greater than $X_{sup}$ imply negative $c_1$ contradicting its definition. The continuity together with asymptotic behavior of the critical curve
\begin{align}
	\lim_{\X \rightarrow 0} \left( c_1(\X), c_2(\X) \right) &= (\infty, 0) \\
    \lim_{\X \rightarrow \X_{sup}} \left( c_1(\X), c_2(\X) \right) &= (0, \infty)
\end{align}
indicates that it separates the phase diagram in exactly two regions. Indeed, to have more than two regions, the boundary curve should intersect with itself at some point, which is not possible since both of its components are strictly monotonous functions of $\X$.

\subsubsection{Double negative feedback loop}

The critical curve \ref{Boundary curve double negative feedback} is well-defined and continuous for
\begin{align}
	\X_{inf} < \X < +\infty,
\end{align}
with $\X_{inf} = (m n - 1)^{-\frac{1}{n}}$. As in the case for the double positive feedback loop, the critical curve for the double negative feedback loop separates its phase diagram in exactly two regions. The only way to have more region is if the curve intersects with itself, but this is impossible, as shown by the following proposition.
\begin{proposition}
The critical curve of the double positive feedback \eqref{Boundary curve double negative feedback} does not intersect with itself.
\begin{proof}
A sufficient condition for the absence of self intersection is the strict increase of the slope of the boundary curve $d c_2(\X)/ d c_1(\X)$ with increasing $\X$. Due to the smoothness of both $c_1(\X)$ and $c_2(\X)$, we can write
\begin{equation}
	\frac{d c_2(\X)}{d c_1(\X)} = \frac{d c_2(\X)}{d \X} \left(\frac{d c_1(\X)}{d \X}\right)^{-1} = m \frac{(\X^n + 1)^{\frac{1}{m}}}{\X} \left[(mn -1) \X^n - 1 \right]^{1 - \frac{1}{m}}.
\end{equation}

We now take the derivative with respect to $A$ to find
\begin{equation}
	\label{Derivative of the slope of the boundary curve}
	sign\left(\frac{d}{d \X} \frac{d c_2(\X)}{d c_1(\X)}\right) = sign\left( (n - 1) (m n - 1) \X^{2n} +(n - 1) (mn - n - 2) \X^n + 1 \right).
\end{equation}
This expression can be interpreted as a parabola in terms of $\X^n$. Since the leading coefficient is positive, the quadratic expression yields positive values except in between its two roots $\X^n_-$ and $\X^n_+$. For their values we find
\begin{equation}
	\X^n_{\pm} = \frac{n}{2} \frac{m-1}{mn-1}\left[ \pm \sqrt{1 - \frac{4}{(m-1)(n-1)}} - \left(1 - \frac{2}{n(m-1)}\right)\right].
\end{equation}
From this result, we can verify that $\X_{\pm}^m$ are either negative or complex. As a direct consequence, since $\X$ is always positive, the sign of the derivative of the slope \eqref{Derivative of the slope of the boundary curve} is always positive as well and thus the slope of the boundary curve is strictly increasing with respect to $\X$.
\end{proof}
\end{proposition}
\newpage
\bibliographystyle{siam}
\thispagestyle{empty}
\bibliography{biblio}
\newpage
\appendix
\appendixpage
\section{From chemical reaction network to system of ODEs}
\label{appendix: from chemical to ode}
This section provides a precise method for the transition from a chemical reaction network (Equations \eqref{chemical reaction network double positive} and \eqref{chemical reaction network double negative}) to a system at equilibrium (Equations \eqref{Double positive feedback simplified} and \eqref{Double negative feedback simplified}). This method, presented here for the double positive and double negative feedback loops, can be generalized to other chemical reaction networks.
\subsection{Double positive feedback loop}
The changes over time in the concentrations of the factors $A$ and $B$ in the positive feedback loop network \eqref{chemical reaction network double positive} (Figure \ref{double.1}), and in the subsequent multimerization complexes $nA$ and $mB$, can be described by the mass action law, leading to the following differential equations:

\begin{equation}
    \begin{alignedat}{2}
\label{Mass action equations}
	&\frac{d}{dt}c_{\mathcal{O}^B} &&= -\kappa^B_f (c_{A}^n) c_{\mathcal{O}^B}+\kappa^B_b c_{\mathcal{O}^B_{nA}}\\
	&\frac{d}{dt}c_{\mathcal{O}^B_{nA}} &&= -\frac{d}{dt}c_{\mathcal{O}^B} \\
    &\frac{d}{dt}c_{A} &&= \mu^Ac_{\mathcal{O}^A_{mB}}-\nu^Ac_{A} + n  \frac{d}{dt}c_{\mathcal{O}^B} \\
    &\frac{d}{dt}c_{\mathcal{O}^A} &&= -\kappa^A_f (c_{B}^m) c_{\mathcal{O}^A}+\kappa^A_bc_{\mathcal{O}^A_{mB}}\\
    &\frac{d}{dt}c_{\mathcal{O}^A_{mB}} &&= -\frac{d}{dt}c_{\mathcal{O}^A} \\
    &\frac{d}{dt}c_B &&= \mu^Bc_{\mathcal{O}^B_{nA}}-\nu^Bc_{B} + m\frac{d}{dt}c_{\mathcal{O}^A},
\end{alignedat}
\end{equation}
where $c$ indicates the concentration of the respective factor or operator. Under chemical equilibrium (no net change in the concentration of any species), the system reduces to:
\begin{align}
\label{Equilibrium stochastic}
\begin{aligned}
\textbf{for factor A}&\\
    c_{A}^n c_{\mathcal{O}^B} - K_B c_{\mathcal{O}^B_{nA}} &= 0\\
    \mu^A c_{\mathcal{O}^A_{mB}}-\nu^Ac_{A} &= 0
\end{aligned}
&&
\begin{aligned}
\textbf{for factor B}&\\
    c_{B}^m c_{\mathcal{O}^A} - K_A c_{\mathcal{O}^A_{mB}} &= 0\\
    \mu^B c_{\mathcal{O}^B_{nA}}-\nu^B c_{B} &= 0
\end{aligned}
\end{align}
with $K_A = \frac{\kappa_b^A}{\kappa_f^A}$ and $K_B=\frac{\kappa_b^B}{\kappa_f^B}$. One can further notice from Equation \eqref{Mass action equations} that the sum of bound and unbound operators is constant, and we can thus define $c_{\mathcal{O}^A_{tot}}$ and $c_{\mathcal{O}^B_{tot}}$ as
\begin{align}
\begin{split}
\label{Total_pos}
c_{\mathcal{O}^A_{tot}} &= c_{\mathcal{O}^A}+c_{\mathcal{O}^A_{mB}}\\
c_{\mathcal{O}^B_{tot}} &= c_{\mathcal{O}^B}+c_{\mathcal{O}^B_{nA}}.
\end{split}
\end{align}
Putting Equation \eqref{Total_pos} into Equation \eqref{Equilibrium stochastic}, gives the fraction of occupied operators
\begin{align}
\label{Occupied_operators}
\begin{split}
    \frac{c_{\mathcal{O}^A_{mB}}}{c_{\mathcal{O}^A_{tot}}} &= \frac{c_{B}^m}{K_A+c_{B}^m} \\
    \frac{c_{\mathcal{O}^B_{nA}}}{c_{\mathcal{O}^B_{tot}}} &= \frac{c_{A}^n}{K_B+c_{A}^n}.
\end{split}
\end{align}
This allows to eliminate the concentrations of bound and unbound operators from the system in Equation (\ref{Equilibrium stochastic}) and to rewrite the relations as
\begin{align*}
\begin{aligned}
\textbf{for factor A}&\\
    \frac{\mu^A c_{\mathcal{O}^A_{tot}} c_{B}^m}{K_A+c_{B}^m}-\nu^A c_{A} &= 0
\end{aligned}
&&
\begin{aligned}
\textbf{for factor B}&\\
    \frac{\mu^B c_{\mathcal{O}^B_{tot}} c_{A}^n}{K_B+c_{A}^n}-\nu^B c_{B} &= 0.
\end{aligned}
\end{align*}
\noindent
Then, by defining the normalized concentrations $\X$ and $\Y$ as
\begin{align}
	\X = \frac{c_{A}}{\sqrt[n]{K_B}}, \qquad \Y = \frac{c_{B}}{\sqrt[m]{K_A}}
	\label{Normalized parameters}
\end{align}
and including the constants $c_1$ and $c_2$
\begin{align}
\label{c1c2pos}
	c_1 = \frac{\mu^A}{\nu^A} \frac{c_{\mathcal{O}^A_{tot}}}{\sqrt[n]{K_B}}, \qquad c_2 = \frac{\mu^B}{\nu^B} \frac{c_{\mathcal{O}^B_{tot}}}{\sqrt[m]{K_A}}
\end{align}
the steady states of the double positive feedback can be written in a simpler form as
\begin{align}
  \begin{cases}
  \dfrac{c_1 \Y^m}{1 + \Y^m} - \X = 0
  \\
  \\
  \dfrac{c_2 \X^n}{1 + \X^n} - \Y = 0.
\end{cases}
\end{align}

\subsection{Double negative feedback loop}
The changes over time in the concentrations of the factors $A$ and $B$ in the negative feedback loop of network \eqref{chemical reaction network double negative} (Figure \ref{double.1}), and in the subsequent multimerization complexes $nA$ and $mB$ can be described by the following differential equations:
\begin{equation}
\begin{alignedat}{2}
\label{Mass action equations negative}
	&\frac{d}{dt}c_{\mathcal{O}^B} &&= -\kappa^B_f (c_{A}^n) c_{\mathcal{O}^B}+\kappa^B_b c_{\mathcal{O}^B_{nA}}\\
	&\frac{d}{dt}c_{\mathcal{O}^B_{nA}} &&= -\frac{d}{dt}c_{\mathcal{O}^B}\\
    &\frac{d}{dt}c_{A} &&= \mu^A c_{\mathcal{O}^A}-\nu^A c_{A} + n\frac{d}{dt}c_{\mathcal{O}^B}\\
    &\frac{d}{dt}c_{\mathcal{O}^A} &&= -\kappa^A_f (c_{B}^m) c_{\mathcal{O}^A}+\kappa^A_b c_{\mathcal{O}^A_{mB}}\\
    &\frac{d}{dt}c_{\mathcal{O}^A_{mB}} &&= -\frac{d}{dt}c_{\mathcal{O}^A} \\
    &\frac{d}{dt}c_{B} &&= \mu^B c_{\mathcal{O}^B}-\nu^B c_{B} + m\frac{d}{dt}c_{\mathcal{O}^A}.
\end{alignedat}
\end{equation}
Under chemical equilibrium, the system reduces to:
\begin{align}
\label{Equilibrium stochastic negative}
\begin{aligned}
\textbf{for factor A}&\\
    c_{A}^n c_{\mathcal{O}^B} - K_B c_{\mathcal{O}^B_{nA}} &= 0\\
    \mu^A c_{\mathcal{O}^A}-\nu^A c_{A} &= 0
\end{aligned}
&&
\begin{aligned}
\textbf{for factor B}&\\
    c_{B}^m c_{\mathcal{O}^A} - K_A c_{\mathcal{O}^A_{mB}} &= 0\\
    \mu^B c_{\mathcal{O}^B}-\nu^B c_{B} &= 0
\end{aligned}
\end{align}
with $K_A = \frac{\kappa_b^A}{\kappa_f^A}$ and $K_B=\frac{\kappa_b^B}{\kappa_f^B}$.
\\

We see that $c_{\mathcal{O}^A_{tot}}$ and $c_{\mathcal{O}^B_{tot}}$ remain constant as in Equation \eqref{Total_pos}, substituting into Equation \eqref{Equilibrium stochastic negative} yields the fraction of free operators
\begin{align}
\label{Free_operators}
\begin{split}
    \frac{c_{\mathcal{O}^A}}{c_{\mathcal{O}^A_{tot}}} &= \frac{K_A}{K_A+c_{B}^m} \\
    \frac{c_{\mathcal{O}^B}}{c_{\mathcal{O}^B_{tot}}} &= \frac{K_B}{K_B+c_{A}^n}.
\end{split}
\end{align}
Substituting this in Equation \eqref{Equilibrium stochastic negative}, we obtain
\begin{align*}
\begin{aligned}
\textbf{for factor A}&\\
    \frac{\mu^A c_{\mathcal{O}^A_{tot}}K_A}{K_A+c_{B}^m}-\nu^Ac_{A} &= 0
\end{aligned}
&&
\begin{aligned}
\textbf{for factor B}&\\
    \frac{\mu^B c_{\mathcal{O}^B_{tot}}K_B}{K_B+c_{A}^n}-\nu^Bc_{B} &= 0.
\end{aligned}
\end{align*}
Defining the normalized concentrations $\X$ and $\Y$ as
\begin{align}
\label{XYneg}
	\X = \frac{c_{A}}{\sqrt[m]{K_B}}, \qquad \Y = \frac{c_{B}}{\sqrt[n]{K_A}}
\end{align}
and the constants $c_1$ and $c_2$ as:
\begin{align}
\label{c1c2neg}
	c_1 = \frac{\mu^A c_{\mathcal{O}^A_{tot}}K_A}{\nu^A}, \qquad c_2 = \frac{\mu^B c_{\mathcal{O}^B_{tot}}K_B}{\nu^B}
\end{align}
we obtain a simpler description of the steady states of the double negative feedback as:
\begin{align}
  \begin{cases}
  \dfrac{c_1}{1 + \Y^m} - \X = 0
  \\
  \\
  \dfrac{c_2}{1 + \X^n} - \Y = 0.
\end{cases}
\end{align}

\section{Rigorous numerical verification}
\label{double:simulation}
In order to further verify this result numerically, we used interval analysis theory described below. Its power resides in the fact it can guarantee that a given system has two stable solutions \cite{Moore2009, Tucker2011}, by first finding a solution and then verifying that the criterion $\delta < 0$ holds. We preferred using interval arithmetic here instead of standard numerical methods, since the latter can in general not guarantee the existence of a solution, nor determine the sign of $\delta$ due to numerical inaccuracies.
\subsection{Interval arithmetic}
Consider a closed interval $X = [a,b]$ defined by 
$$[a,b] =  \lbrace x \in \mathbb{R}:a\leq x \leq b\rbrace.$$
Let $X = [a_1,b_1]$ and $Y=[a_2,b_2]$ be two intervals and $\ast$ an operation such as addition or multiplication for example. The operation $X \ast Y$ between the two intervals $X$ and $Y$ is defined as $$X\ast Y = \lbrace x\ast y:x\in X,y\in Y\rbrace. $$
Operations between intervals can be calculated with the Julia package \emph{IntervalArithmetic.jl} \cite{intervalarithmetic}.
\begin{example}
The four basic operations: addition, subtraction, multiplication and division applied to intervals give the following results
\begin{equation*}
\begin{alignedat}{2}
	&[-3,2]+[1,5] && = [-2,7]\\
	&[-3,2]-[1,5] && =[-8,1]\\
	&[-3,2]*[1,5]&&=[-15,10]\\
	&[-3,2]:[1,5]&&=[-3,2].
\end{alignedat}
\end{equation*}
\end{example}
More complicated concepts like functions or matrices can also be applied to intervals, for more information, the reader can refer to \cite{Moore2009}.

\subsection{The Krawczyk operator}

In order to solve a system of $m$ nonlinear equations in $\mathbb{R}^n$:
\begin{align}
\begin{cases}
	f_1(x_1,...,x_n) &= 0\\
    \hspace{0.5cm}\vdots\\
	f_m(x_1,...,x_n) &= 0,
\end{cases}
\end{align}
this system can be written as
\begin{equation}
\label{krawczyk f}
f(x)= 0
\end{equation}
with $f(x) = (f_1(x),...,f_m(x))$ and $x=(x_1,...,x_n)$.
\\

We next introduce the Krawczyk operator.
\begin{definition}[The Krawczyk operator]
\emph{The Krawczyk operator} is defined as follows
$$K(X) = y-Yf(y)+(I-YF'(X))(X-y)$$
where $X$ is an interval that contains the starting point $y\in\mathbb{R}^n$, $Y=(f'(y))^{-1}$, $I$ the identity matrix and $F'(X)$ an interval extension of the Jacobian matrix. 
\end{definition}
The Krawczyk operator is computed by the Julia package \emph{IntervalRootFinding.jl} \cite{intervalrootfinding}. This package uses the following theorem to find the solutions of a system of nonlinear equations.
\begin{theorem}[\cite{Moore2009}]
\label{th moore}
Let $D = [a_1,b_1]\times ... \times [a_n,b_n]$ be an interval box.
\begin{enumerate}
    \item If 
    \begin{equation}
        K(D) \subset D,
        \label{thmoore_1}
    \end{equation}
     then the system \eqref{krawczyk f} has a unique root in $D$,
    \item If 
    \begin{equation}
        K(D)\cap D =\emptyset,
        \label{thmoore_2}
    \end{equation}
    then the system \eqref{krawczyk f} has no root in $D$.
\end{enumerate}
\end{theorem}

\subsection{Algorithm 1}
The following algorithm is used by \emph{IntervalRootFinding.jl} to compute all the zeros of a system $f(x)=0$ in an interval $D_0= [a_1,b_1]\times ... \times [a_n,b_n]$. The different steps of the algorithm are the followings:
\begin{enumerate}[label= {\emph{Step \arabic*.}}, start=0, leftmargin=1.5cm, rightmargin=1cm]
    \item Set a tolerance level $t$.
    \item Add $D_0$ to the list of unprocessed intervals.
    \item Retrieve one of the unprocessed intervals and name it $D$. If there is no interval left to process terminate.
    \item Check the condition \eqref{thmoore_1} for $D$. If it is fulfilled store $K(D)$ as an interval containing a unique solution and go back to \textit{Step 2}.
    \item Check the condition \eqref{thmoore_2} for $D$. If it is fulfilled, discard $D$ and go back to \textit{Step 2}.
    \item If $rad(D) < t$ go back to \textit{Step 2}. Otherwise, bisect $D$ in two intervals $D_1$ and $D_2$. Add $D_1$ and $D_2$ to the list of unprocessed intervals and go back to \textit{Step 2}.
\end{enumerate}
\subsubsection{Choice of the interval $D_0$}
To apply Theorem \ref{th moore} to Equations \eqref{Double positive feedback simplified} and \eqref{Double negative feedback simplified} it is necessary to define a starting interval $D_0= [x_1,x_2]\times [y_1,y_2]$ which must be as small as possible for the algorithm to be efficient but which must also be large enough to contain all solutions of the system. It is natural to choose $x_1 = y_1 = 0$ as a lower bound for the two intervals $[x_1,x_2]$ and $[y_1,y_2]$ because the variables $X$ and $Y$ represent concentrations and are therefore positive. For the upper bounds, we fix the two parameters $c_1$ and $c_2$ and transform Equations \eqref{Double positive feedback simplified} and \eqref{Double negative feedback simplified} as follows:
\begin{equation}
    \label{sharppos}
\begin{cases}
\dfrac{c_1Y^m}{1+Y^m} - X = 0
\\
\\
\dfrac{c_2X^n}{1+X^n} - Y = 0
\end{cases}
\Leftrightarrow
\begin{cases}
\dfrac{c_1}{\frac{1}{Y^n}+1} = X
\\
\\
\dfrac{c_2}{\frac{1}{X^n}+1} = Y
\end{cases}
\end{equation}
and
\begin{equation}
    \label{sharpneg}
\begin{cases}
\dfrac{c_1}{1+Y^m} - X = 0
\\
\\
\dfrac{c_2}{1+X^n} - Y = 0
\end{cases}
\Leftrightarrow
\begin{cases}
\dfrac{c_1}{1+Y^m} = X
\\
\\
\dfrac{c_2}{1+X^n} = Y,
\end{cases}
\end{equation}
which implies $\X\leq c_1$ and $\Y\leq c_2$.
\\

To see the sharpness of the upper bounds $c_1$ and $c_2$, we let $n\longrightarrow\infty$ and $m\longrightarrow\infty$, and we obtain from Equations \ref{sharppos} and \ref{sharpneg}:
\begin{equation*}
    \begin{cases}
    \X \xlongrightarrow[n\longrightarrow\infty]{\Y>1} c_1
    \\
    \\
    Y \overset{c_1 > 1}{\longrightarrow} c_2
    \end{cases}
    \text{and }
    \begin{cases}
    X \overset{Y<1}{\longrightarrow} 0
    \\
    \\
    Y \overset{}{\longrightarrow} 0
    \end{cases}
    \end{equation*}
    and
    \begin{equation*}
    \begin{cases}
    X \overset{Y<1}{\longrightarrow} c_1
    \\
    \\
    Y \overset{c_1 > 1}{\longrightarrow} 0
    \end{cases}
    \text{and }
    \begin{cases}
    X \overset{Y>1}{\longrightarrow} 0
    \\
    \\
    Y \overset{}{\longrightarrow} c_2.
    \end{cases}
    \end{equation*}
The upper bound $c_1$ for the interval $X$ and the upper bound $c_2$ for the interval $Y$ are therefore a judicious choice for $x_2$ and $y_2$ respectively. We thus obtain $$D_0 = [0,c_1]\times [0,c_2]$$ as the interval for initializing the simulations.

\subsection{Algorithm 2}
The following algorithm, developed around Theorem \ref{th moore}, has been used to numerically verify the bistability of the systems \eqref{Double positive with feedback rewritten} and \eqref{Double negative with feedback rewritten} as a function of the parameters $c_1$ and $c_2$. Here are the different steps of this algorithm:
\begin{enumerate}[label= {\emph{Step \arabic*.}}, start=0, leftmargin=1.5cm, rightmargin=1cm]
	\item Set initial values $c_{1_{max}}$ and $c_{2_{max}}$ for the maximal values for $c_1$ and $c_2$ respectively.
	Set a step $s$, a tolerance level $t$ and an empty set $S$.
	\item Define two sets
	\begin{equation}
	    \begin{split}
	        L_{c_1} &= \big\{k s : k\in\mathbb{N}, k s < c_{1_{max}} \big\}\\
	        L_{c_2} &= \big\{k s : k\in\mathbb{N}, k s < c_{2_{max}} \big\}
	    \end{split}
	\end{equation}
	\item For $c_1\in L_{c_1}$ and $c_2\in L_{c_2}$, define the interval
	\begin{equation}
	    D=[0,c_1]\times[0,c_2]
	\end{equation}
	\item Use \emph{Algorithm 1} on the interval $D$. If the algorithm finds three distinct solutions, stores the configuration $\{c_1,c_2\}$ in the set $S$. Repeat from \textit{Step 2} for all $c_1\in L_{c_1}$ and $c_2\in L_{c_2}$.
\end{enumerate}
Since having three steady states implies to have two stable steady states, at the end of the algorithm the set $S$ will contain all the configuration $\{c_1,c_2\}$ for which the system is bistable. Figure \ref{Figure5.7} shows the results of the numerical verification.
\end{document}